%% file: prexl.tex
\documentclass[twocolumn,showpacs,preprintnumbers,amsmath,amssymb,prl,
superscriptaddress]{revtex4}

\usepackage{graphicx}
\usepackage{dcolumn}
\usepackage{bm}
\usepackage{color}

\input def.sty

\begin{document}

\title{Measurement of the Neutron Radius of  $^{208}$Pb Through Parity-Violation in Electron Scattering}

\input author-list_final.tex

\date{\today}

\begin{abstract}
We report the first measurement of the parity-violating asymmetry 
$\APV$ in the elastic scattering of polarized electrons from $^{208}$Pb.
$\APV$ is sensitive to the radius of the neutron distribution $(R_n)$. The result 
$\APV=0.656\pm0.060\ {\rm (stat)}\pm0.014\ {\rm (syst)}$ ppm corresponds
to a difference between the radii of the neutron and proton distributions $R_n-R_p= 0.33{\scriptsize\begin{array}{c}
{+0.16}\\{-0.18}\end{array}}$ fm and provides the first electroweak observation of the neutron
skin which is expected in a heavy, neutron-rich nucleus.
\end{abstract}

\pacs{21.10.Gv, 21.65.Ef, 25.30.Bf,  27.80.+w $190\le A \le  219$}

\maketitle

\input Pintro.tex

\input Panalysis

\input ack

\input Pbib
\end{document}

%% file: author-list_final.tex
\affiliation{Argonne National Laboratory, Argonne, Illinois 60439, USA} 
\affiliation{California State University, Los Angeles, Los Angeles, California 90032, USA} 
\affiliation{Carnegie Mellon University, Pittsburgh, Pennsylvania 15213, USA} 
\affiliation{Clermont Universit\'e, Universit\'e Blaise Pascal, CNRS/IN2P3,
Laboratoire de Physique Corpusculaire, FR-63000 Clermont-Ferrand, France}
\affiliation{College of William and Mary, Williamsburg, Virginia 23187, USA} 
\affiliation{Christopher Newport University, Newport News, Virginia 23606, USA} 
\affiliation{Florida International University, Miami, Florida 33199, USA} 
\affiliation{Hampton University, Hampton, Virginia 23668, USA} 
\affiliation{Harvard University, Cambridge, Massachusetts 02138, USA} 
\affiliation{INFN, Sezione di Bari and University of Bari, I-70126 Bari, Italy} 
\affiliation{INFN, Dipt.~di Fisica dell'Univ.~di Catania, I-95123 Catania, Italy} 
\affiliation{INFN, Sezione di Roma, I-00161 Rome, Italy} 
\affiliation{INFN, Sezione di Roma, gruppo Sanit\`a, I-00161 Rome, Italy} 
\affiliation{Indiana University, Bloomington, Indiana 47405, USA} 
\affiliation{Institut Jo\v zef Stefan, 3000 SI-1001 Ljubljana, Slovenia}
\affiliation{Kharkov Institute of Physics and Technology, Kharkov 61108, Ukraine} 
\affiliation{Laboratoire de Physique Corpusculaire, Clermont-Ferrand Campus des Cézeaux, 63171 Aubière Cedex, France} 
\affiliation{Lawrence Berkeley National Laboratory, Berkelery, California 94720, USA} 
\affiliation{Longwood University, Farmville, Virginia 23909, USA} 
\affiliation{Los Alamos National Laboratory, Los Alamos, New Mexico 87545, USA} 
\affiliation{Massachusetts Institute of Technology, Cambridge, Massachusetts 02139, USA} 
\affiliation{Mississippi State University, Mississippi State, Mississippi 39762, USA} 
\affiliation{Ohio University, Athens, Ohio 45701, USA} 
\affiliation{Old Dominion University, Norfolk, Virginia 23529, USA} 
\affiliation{Rensselaer Polytechnic Institute, Troy, New York 12180, USA} 
\affiliation{Rutgers University, The State University of New Jersey, New Brunswick, New Jersey 08901, USA} 
\affiliation{Seoul National University, Seoul 151-742, South Korea} 
\affiliation{Smith College, Northampton, Massachusetts 01063, USA} 
\affiliation{Syracuse University, Syracuse, New York 13244, USA} 
\affiliation{Tel Aviv University, P.O. Box 39040, Tel-Aviv 69978, Israel} 
\affiliation{Temple University, Philadelphia, Pennsylvania  19122, USA} 
\affiliation{Thomas Jefferson National Accelerator Facility, Newport News, Virginia 23606, USA}  
\affiliation{University of Massachusetts Amherst, Amherst, Massachusetts  01003, USA} 
\affiliation{University of New Hampshire, Durham, New Hampshire 03824, USA} 
\affiliation{University of Science and Technology of China, Hefei, Anhui 230026, P.R. China} 
\affiliation{University of Virginia, Charlottesville, Virginia  22903, USA} 
\affiliation{Virginia Polytechnic Institute and State University, Blacksburg, Virginia  24061, USA} 
\affiliation{Yerevan Physics Institute, Yerevan, Armenia} 
\author{S.~Abrahamyan}\affiliation{Yerevan Physics Institute, Yerevan, Armenia} 
\author{Z.~Ahmed}\affiliation{Syracuse University, Syracuse, New York 13244, USA} 
\author{H.~Albataineh}\affiliation{Laboratoire de Physique Corpusculaire, Clermont-Ferrand Campus des Cézeaux, 63171 Aubière Cedex, France} 
\author{K.~Aniol}\affiliation{California State University, Los Angeles, Los Angeles, California  90032, USA} 
\author{D.~S.~Armstrong}\affiliation{College of William and Mary, Williamsburg, Virginia  23187, USA} 
\author{W.~Armstrong}\affiliation{Temple University, Philadelphia, Pennsylvania  19122, USA} 
\author{T.~Averett}\affiliation{College of William and Mary, Williamsburg, Virginia  23187, USA} 
\author{B.~Babineau}\affiliation{Longwood University, Farmville, Virginia 23909, USA} 
\author{A.~Barbieri}\affiliation{University of Virginia, Charlottesville, Virginia  22903, USA} 
\author{V.~Bellini}\affiliation{INFN, Dipt.~di Fisica dell'Univ.~di Catania, I-95123 Catania, Italy} 
% ...
\collaboration{PREX Collaboration}

%% file: Pintro.tex

Nuclear charge densities have been accurately measured with electron scattering
and have become our picture of the atomic nucleus, see for 
example~\cite{chargeden}.  
In contrast, our knowledge of neutron densities comes primarily from hadron
scattering experiments involving, for example, pions \cite{pions}, protons
\cite{protons1,protons2,protons3}, or antiprotons
\cite{antiprotons1,antiprotons2}, the interpretation of which requires
a model-dependent description of the non-perturbative strong interaction.
Due to the fact that the weak charge of the neutron is much larger than that of the proton, 
the measurement of parity violation in electron scattering provides a model-independent probe of
neutron densities that is free from most strong-interaction uncertainties~\cite{dds}.

In the Born approximation, the parity violating cross-section asymmetry for
longitudinally polarized electrons elastically scattered from an unpolarized nucleus,  $A_{PV}$,
is proportional to the weak form factor $F_W(Q^2)$.  This is the Fourier transform of the weak
charge density, which is closely related to the neutron density, and therefore the neutron
density can be
extracted from an electro-weak measurement \cite{dds}.
\begin{equation}
A_{PV}=\frac{\sigma_R-\sigma_L}{\sigma_R+\sigma_L} \approx \frac{G_FQ^2}{4\pi\alpha\sqrt{2}}
\frac{F_W(Q^2)}{F_{ch}(Q^2)}
\end{equation}
where $\sigma_{R(L)}$ is the differential cross section for elastic scattering of right (R) and
left (L) handed longitudinally polarized electrons, $G_F$ is the Fermi constant, $\alpha$ the
fine structure constant, and $F_{ch}(Q^2)$ is the Fourier transform of the known charge density.
However, the Born
approximation is not valid for a heavy nucleus and Coulomb-distortion effects
must be included.  These have been accurately calculated \cite{couldist}
because the charge density is well known, and many
other details relevant for a practical parity-violation experiment to 
measure neutron densities have been 
discussed in a previous publication~\cite{bigprex}.    


One system of particular interest is the doubly-magic nucleus $^{208}$Pb, which
has 44 more neutrons than protons; some
of these extra neutrons are expected to be found in the surface, where they form
a neutron-rich skin.  The thickness of this skin is sensitive to nuclear
dynamics and provides fundamental nuclear structure information.  
A number of mean-field-theory models have been developed that
agree with the world's body of data on nuclear charge distributions and other nuclear 
properties~\cite{Lalazissis:1996rd,ToddRutel:2005zz,Beiner:1974gc,Chabanat:1997un,Vautherin:1971aw}.  
For $\pb$, these are consistent with a radius of the point-neutron distribution $R_n$ between 0.0 -- 0.4~fm 
larger than that of the point-proton distribution $R_p$.  In this paper we report a first measurement of $\APV$ from $\pb$, 
which is sensitive to the existence of the neutron skin. 

The value of the neutron radius of $\pb$ has important implications for models of nuclear structure and their 
application in atomic physics and astrophysics. 
There is a strong correlation between $R_n$ of $\pb$ and the pressure of
neutron matter $P$ at densities near 0.1 fm$^{-3}$ (about 2/3 of nuclear
density) \cite{alexbrown}.  A larger $P$ will push neutrons out against surface
tension and increase $R_n$.  Therefore measuring $R_n$ constrains the 
equation of state (EOS), the pressure as a function of density, of neutron matter.  

The correlation between $R_n$ and the radius of a neutron star $r_{NS}$ is also
very interesting \cite{rNSvsRn}.  In general, a larger $R_n$ implies a stiffer
EOS, with a larger pressure, that will also suggest $r_{NS}$ is larger.  
Recently there has been great progress in deducing $r_{NS}$ from X-ray
observations. From observations of X-ray bursts, $\mbox{Ozel {\it et al}.\ \cite{Ozel:2010fw}}$ find $r_{NS}$ is very 
small, near 10~km,  implying that the EOS softens at high density which is suggestive of a 
transition to an exotic phase of QCD.  In contrast,  $\mbox{Steiner {\it et al}.\ \cite{Steiner:2010fz}}$ conclude
that $r_{NS}$ is near 12 km, leading to a prediction that $R_n-R_p=0.15 \pm 0.02$~fm for $\pb$. This implies a stiffer
EOS which leaves little room for softening due to a phase transition at high density.

Recently $\mbox{Hebeler {\it et al}.\ \cite{Hebeler:2010jx}}$ used chiral perturbation theory to
calculate the EOS of neutron matter including important contributions 
from 
three-neutron forces.  
From their EOS, they predict $R_n-R_p= 0.17 \pm 0.03$~fm for $\pb$.
Monte Carlo calculations by $\mbox{Carlson {\it et al}.\ \cite{MC3n}}$ also find sensitivity to three-neutron forces.   
The measurement of $R_n$ provides an important check of fundamental neutron matter 
calculations, and constrains three-neutron forces.  

The EOS of neutron-rich matter is closely related to the symmetry energy $S$.  
There is a strong correlation between $R_n$ and the density dependence of the
symmetry energy $dS/d\rho$, with $\rho$ as the baryon density.  The symmetry
energy can be probed in heavy-ion collisions \cite{isospindif}.  For example,
$dS/d\rho$ has been extracted from isospin diffusion data \cite{isospindif2} 
using a transport model.

The symmetry energy $S$ helps determine the composition of a neutron star.
A large $S$ at high density would imply a large proton fraction,
which would allow the direct Urca process \cite{URCA} of rapid neutrino cooling. 
If $R_n-R_p$ in $\pb$ were
large, it is likely that massive neutron stars would cool quickly by direct Urca.
In addition, the transition density from a solid neutron star crust
to the liquid interior is strongly correlated with $R_n-R_p$ \cite{cjhjp_prl}.  

Reinhard and Nazarewicz claim that $R_n-R_p$ is tightly correlated with the dipole 
polarizability $\alpha_D$ ~\cite{Reinhard} 
and Tamii et al. use this correlation to infer $R_n-R_p$ from 
a new measurement of $\alpha_D$ ~\cite{tamii}.

Atomic parity violation (APV) is also sensitive to 
$R_n$~\cite{bigprex,Pollock,brownAPV}.  A future low-energy test of
the standard model may involve the combination of a precise APV experiment
along with PV electron scattering to constrain $R_n$ ~\cite{Pollock}.
Alternatively, measuring
APV for a range of isotopes could provide information 
on neutron densities \cite{berkeleyAPV}.

%% file: Panalysis.tex
The measurement 
was carried out in Hall A at  the Thomas Jefferson National Accelerator Facility. 
The experimental configuration is similar to that used previously
for studies of the weak form factor of the proton and $^4$He~\cite{Acha:2006my,Aniol:2005zf,Aniol:PRC2004}.
A 50 to 70~$\mu$A continuous-wave beam of longitudinally polarized 1.06~GeV electrons
was incident on a 0.55~mm thick isotopically pure $^{208}$Pb target foil. 
A $4~\mbox{mm}\times4~\mbox{mm}$ square beam raster prevented the target from melting. 
Two 150~$\mu$m diamond foils sandwiched the lead foil to improve thermal conductance to a copper 
frame cooled to 20K with cryogenic helium.  Elastically 
scattered electrons
were focused onto thin quartz detectors in the twin 
High Resolution Spectrometers (HRS)~\cite{Alcorn:2004sb}.
The addition of a pair of dipole septum magnets between the 
target and the HRSs allowed us to achieve a forward scattering angle of $\theta_{\mathrm lab}\sim 5^{\circ}$.
The HRS momentum resolution 
ensured that only elastic events (and a negligible fraction of inelastic events from the 2.6 MeV first excited state) 
were accepted by the quartz detectors.
Cherenkov light from each quartz bar traversed air  
light guides and were detected by 2-inch quartz-window photo-multipliers (PMT).

The polarized electron beam originated from a strained GaAsP 
photocathode illuminated by 
circularly polarized light \cite{Sinclair2007}.
The accelerated beam was directed
into Hall A, where its intensity, energy, polarization,
and trajectory on target were inferred
from the response of several monitoring devices.
The sign of the laser circular polarization determined the electron helicity; this 
was held constant for periods of 8.33~ms, referred to as ``windows''.  
The integrated responses of detector PMTs and beam monitors
were digitized by an 18-bit ADC and recorded for each window.
Two "window quadruplet" patterns of helicity states ($+--+$ or $-++-$) ensured 
complementary measurements at the same phase relative to the 60~Hz line power, 
thus canceling power-line noise from the asymmetry measurement.
The right-left helicity asymmetry in the integrated detector response,
normalized to the beam intensity,
was computed for sets of complementary helicity windows in each
quadruplet to form the raw asymmetry
$A_{raw}$.   The sequence of these patterns was
chosen with a pseudo-random number generator.  

Loose requirements were imposed on beam quality,
removing periods of position, energy, or beam-intensity instability. 
No helicity-dependent cuts were applied, leaving a final data sample of $2\times10^7$
helicity-window quadruplets.
The design of the apparatus ensured that, after all corrections, 
the fluctuations in the fractional difference of the PMT response between
a pair of successive windows was
dominated by scattered-electron counting statistics for rates up to 1~GHz. This facilitated 
the ability to achieve  an $\APV$ precision significantly better than 100 parts per billion (ppb) in a reasonable length of time. 
Careful attention to the design and configuration of the photocathode laser 
optics~\cite{Paschke:2007zz} ensured that spurious beam-induced asymmetries were under control at this level.

Random fluctuations in beam position and energy contributed the largest source of noise beyond counting statistics in $\ARAW$.
Typical beam jitter in window-quadruplets was 
less than 
2 parts per million (ppm) in energy, and 20 $\mu$m in position.
This noise contribution was reduced by measuring window differences $\Delta x_i$ using 
beam position monitors and applying a correction $A_{beam}=\sum c_i\Delta x_i$.
The $c_i$'s were measured several times each hour
from calibration data in which the beam was modulated
by using steering  coils and an accelerating cavity.
The largest of the $c_i$'s was $\sim$ 50 ppm/$\mu$m.
The noise in the resulting $A_{corr}=A_{raw}-A_{beam}$ was 210~(180)~ppm
per quadruplet, for a beam current of 50~(70)~$\mu$A, dominated by counting statistics 
($\sim$ 1~GHz at 70 $\mu$A). 
Non-uniformities in target thickness due to thermal damage caused  
window-to-window luminosity fluctuations from variations in the target 
area sampled by the rastered beam, leading to the degradation of $\ACORR$ by 
$\sim 40$\%. This source of noise was eliminated by locking the raster
pattern frequency to a multiple of the helicity frequency.  
Low-current calibration data, triggered
on individual scattered electrons, were regularly collected to evaluate the thickness
of lead relative to diamond.

Sensitivity of $\ACORR$ to a transverse component of the beam polarization, coupled to 
the vector analyzing powers ($A_T$) for $^{208}$Pb and $^{12}$C, was studied using special runs with
fully transverse beam polarization. The symmetry of the detector configuration as well 
as the measured $A_{T}$ values (to be published separately) resulted in an upper bound for a 
possible correction to $\ACORR$ of 0.2\%.
The $A_{raw}$ and $A_{corr}$ window-pair distributions for the two 
complete data samples had negligible non-Gaussian tails over more than four 
orders of magnitude. 
To test the accuracy of error calculations and general statistical behavior of the data,
$A_{corr}$ averages and statistical errors were studied for typical one-hour runs,
consisting of $\sim 50$k quadruplets each.
This set of 316 average $A_{corr}$ values, normalized by 
the corresponding statistical error, populated a Gaussian 
distribution of unit variance, as expected.

A half-wave ($\lambda$/2) plate was periodically inserted into the 
injector laser optical path, reversing the
sign of the electron beam polarization relative to both the electronic helicity control signals
and the voltage applied to the polarized source laser electro-optics.
Roughly equal statistics were collected with this waveplate inserted and retracted, suppressing 
many possible sources of systematic error.  An independent method of helicity reversal was
feasible with a pair of Wien spin-rotators separated by a solenoid,  providing an additional powerful
check of systematic control.  
Reversing the direction of the solenoidal field reversed the electron beam helicity 
while the beam optics, 
which depend on the square of the solenoidal magnetic field, were unchanged.  
The $\lambda/2$ reversal was done about every 12 hours and the 
magnetic spin reversal was performed every few days.The dataset consisting of a period between two successive 
$\lambda/2$ or magnetic spin-reversals is referred to as a ``slug''.

The spin reversals resulted in cumulative differences in beam position and energy of only 4 nm and 0.6 ppb respectively, 
leading to a run-averaged $A_{beam} = -39.0 \pm 5.9$ ppb. The asymmetry in beam charge, corrected
by the intensity normalization of $\ARAW$, was $84.0 \pm 1.3$~ppb, with the error determined using 
the correlation of measured beam intensity to PMT response which demonstrated the beam intensity monitors 
were linear to better than 1.5\%. 
Nonlinearity in the PMT response was limited to 1\%\ in bench-tests that mimicked running conditions.
As shown in Table~\ref{table:Acorr}, the values of $A_{corr}$ are consistent within statistical errors 
for each of the reversal states.
The reduced $\chi^2$ for $A_{corr}$ ``slug'' averages
is close to one in every case, indicating that any residual 
beam-related systematic effects were small and randomized over the time 
period of $\lambda/2$ reversals.
The final result is
$A_{corr}= 594\pm 50\rm{(stat)}\pm 9\rm{(syst)}$~ppb where the systematic uncertainty includes possible
effects from $A_{beam}$, non-linearity in the detectors or beam charge monitors, and transverse asymmetry.
The physics asymmetry $A_{PV}$ is formed from $A_{corr}$ by correcting for the
beam polarization $P_b$ and background fractions $f_i$ with asymmetries $A_i$

\begin{equation}
A_{PV} = \frac{1}{P_b}\frac{A_{corr} - 
P_b\sum_{i} A_{i}f_{i}}{1-\sum_{i} f_{i}}.
\end{equation}
These corrections are summarized in
Table~\ref{table:Acorrections}. 

\begin{table}
\begin{tabular}{|c|c|c|c|c|}\hline
$\lambda/2$ plate & Spin-rotator & $A_{corr}$ (ppb) & $\delta A_{corr}$ (ppb)  & $\chi^2 / {\rm d.o.f.}$ \\
\hline
OUT & RIGHT & \ \ 606\ \  & \ \ 113\ \ &  1.03 \\
IN & RIGHT & 492 & 107 & 0.74\\
OUT & LEFT & 565 & 95 & 1.12\\
IN & LEFT & 687 & 92 & 1.03\\
\hline \hline
\multicolumn{2}{|c|}{Average} & 594 & 50 &  $0.99$ \\ \hline
\end{tabular}
\caption{Values of $A_{corr}$ and the statistical error, for each helicity reversal state and for the grand average. The $\chi^2$ per degree of freedom for each average is also shown.}
\label{table:Acorr}
\end{table}


\begin{table}
\begin{tabular}{|l|rcl|rcl|}\hline
Correction & \multicolumn{3}{c|}{Absolute (ppb)} & 
\multicolumn{3}{c|}{Relative(\%)}\\ 
\hline \hline
Beam Charge Normalization & \, -84.0 & $\pm$ & $1.5$  & \, -12.8  & $\pm$ & $0.2$ \\
Beam Asymmetries $A_{beam}$   & 39.0 & $\pm$ & $7.2$  & 5.9  & $\pm$ & $1.1$ \\
Target Backing & $-8.8$   & $\pm$ & $2.6$  & -$1.3$  & $\pm$ & $0.4$ \\
Detector Nonlinearity     & $0$  & $\pm$ & $7.6$   & $0$     & $\pm$ & $1.2$\\
Transverse Asymmetry & $0$   & $\pm$ & $1.2$  & $0$  & $\pm$ & $0.2$ \\
Polarization $P_b$ &  70.9 & $\pm$ & 8.3  & $10.8$ & $\pm$ & $1.3$\\
\hline \hline
Total &  17.1 &$\pm$ & 13.7   & 2.6  & $\pm$ & $2.1$\%\\
\hline \hline
\end{tabular}
\caption{ Corrections to $A_{PV}$ and systematic errors.}
\label{table:Acorrections}
\end{table}

\begin{figure}
\includegraphics[width=1.0\columnwidth]{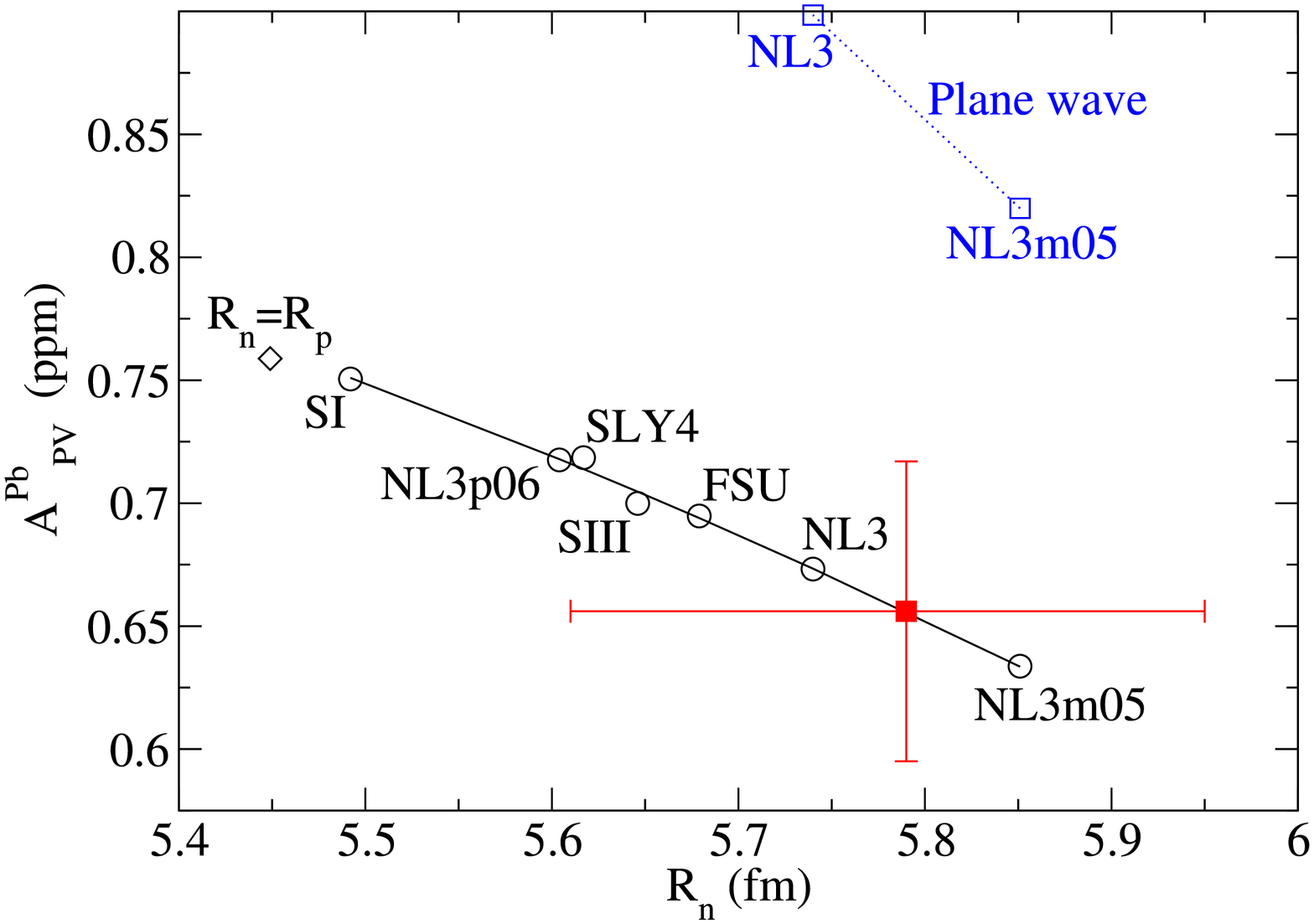}
\caption{Result of this experiment (red square) vs neutron point radius $R_n$ in ${}^{208}$Pb.  
Distorted-wave calculations for seven mean-field neutron densities are 
circles~\cite{Ban:2010wx}, while the diamond marks the expectation for $R_n=R_p$.  The
blue squares show plane wave impulse approximation results.}
\label{fig:R}
\end{figure}

The fraction of the accepted flux from $^{12}$C in the 
detectors varied with time due to changes in the target; averaged over the run, the fraction $f= (6.3 \pm 0.6)\%$.  The asymmetry of this background was determined to be $\APV^{C} = 817 \pm 41$~ppb using the Standard Model  value for the $e$-N weak neutral isoscalar coupling and the measured kinematics, with the uncertainty bounded by the precision measurement of $\APV$ from $^4$He~\cite{Acha:2006my}. This was the only non-negligible background.
An additional possible systematic error in $\langle\qsq\rangle$ lay in the 
determination of the absolute value of $\theta_{lab}$. A
nuclear recoil technique using a water
cell target~\cite{Aniol:2005zf} limited the scale error 
on $\langle\qsq\rangle$ to $1\%$.

The spectrometer acceptance function $\epsilon(\theta)$ characterizes 
the probability, as a function of scattering angle $\theta$, for an electron to 
reach the detector after elastically scattering from $\pb$.
For example, the asymmetry averaged over the acceptance would be
\begin{equation}
\langle A \rangle = \frac{ \int  d\theta  \sin \theta \hskip 0.02in  A(\theta) \hskip 0.02in \frac{d\sigma}{d\Omega} \epsilon(\theta)}
{\int  d\theta \sin \theta \frac{d\sigma}{d\Omega} \epsilon(\theta)}
\end{equation}
where $\frac{d\sigma}{d\Omega}$ is the cross section.
See Supplemental Material at http://hallaweb.jlab.org/parity/prex/accept for the acceptance function $\epsilon(\theta)$.
The observed distribution of events corrected
for the cross section, the background from the carbon (diamond) backing, and the effects of 
multiple scattering is used to extract $\epsilon(\theta)$; 
corrections for energy loss in the target were negligible.  
To compare to predictions, one must integrate the theoretical asymmetry and the $Q^2$
over $\epsilon(\theta)$. 
The systematic error in $\epsilon(\theta)$ was evaluated from reasonable
variations in the parameters of the simulation and resulted in an additional equivalent error in $\langle\qsq\rangle$ of 0.8\%.  
Added in quadrature to the error arising from knowledge of  $\langle\theta\rangle$, we
obtain an overall error in $\langle Q^2 \rangle$ of 1.3\%. We do not include this uncertainty in the total systematic uncertainty of the asymmetry.  Using a calculation by Horowitz~\cite{couldist}, $\mathrm{d}A_{PV}/\mathrm{d}Q^2$ is approximately $30~\mathrm{ppm}/\mathrm{GeV}^2$, which would correspond to an additional systematic uncertainty on $A_{PV}$ of $3~\mathrm{ppb}$ (0.5\% of $A_{PV}$).

The beam polarization was continuously monitored by a Compton polarimeter.
Helicity-dependent asymmetries in the integrated signal from
backscattered Compton 
photons yielded $P_{b} = (88.2 \pm 0.1\pm 1.0)\%$
averaged over the duration of the run.  The beam polarization was stable
within systematic errors. 
An independent M{\o}ller polarimeter making nine measurements at different times 
during the run gave
$P_{b} = (90.3 \pm 0.1\pm 1.1)$\%.
We used an average of these two measurements, $P_{b} = (89.2 \pm 1.0)$\% which conservatively accounts for the correlated systematic errors between the two measurements.

After all corrections,
\[
\APV^{Pb} = 656 \pm 60 \,\,\mbox{(stat)} \,
\pm 14 \,\,\mbox{(syst) ppb}
\]
at $\langle Q^2 \rangle = 0.00880\pm 0.00011\gevc$.
This result is displayed in Figure~\ref{fig:R}, in which models predicting the 
point-neutron radius illustrate the correlation of $\APV^{Pb}$ 
and $R_n$~\cite{Ban:2010wx}. 
For each model, the calculation is performed using the neutron and proton weak
charges $q_n = 0.9878$ and $q_p=-0.0721$
and using the modeled neutron density but the experimental charge density. The importance of Coulomb distortions is emphasized by indicating results from plane-wave calculations, which are not all contained within the vertical axis range of the figure.  A second-order polynomial fit over these models, as illustrated, implies a value for $R_n =5.78{\scriptsize\begin{array}{c}{+0.16}\\{-0.18}\end{array}}$~fm.
Assuming a point-proton radius of 5.45~fm \cite{Ong}, 
corresponding to the measured charge radius of 5.50~fm~\cite{chargeden}, 
implies that the neutron distribution is 1.8$\sigma$ larger than that of
the protons:
$R_n-R_p = 0.33{\scriptsize\begin{array}{c}{+0.16}\\{-0.18}\end{array}}$~fm~\cite{Ban:2010wx} (see
also~\cite{RocaMaza:2011pm}).
A future run is planned which will reduce the quoted uncertainty by a factor of three, to discriminate between models and allow predictions relevant for the description of neutron stars and parity violation in atomic systems.

%% file: ack.tex
\begin{acknowledgments}
We wish to thank the entire staff of JLab for their efforts to develop and
 maintain the polarized beam and the experimental apparatus. 
This work was supported by the U.S. Department of Energy, the National Science Foundation,
and from the French CNRS/IN2P3 and ANR.
Jefferson Science Associates, LLC,  operates Jefferson Lab for the U.S. DOE under U.S. DOE 
contract DE-AC05-060R23177. 

\end{acknowledgments}

%% file: prexl.bbl
\begin{thebibliography}{99} 
\bibitem{chargeden}B. Frois \etal, Phys. Rev. Lett. {\bf 38}, 152 (1977).
\bibitem{pions}C.~Garcia-Recio, J.~Nieves, E.~Oset,
 Nucl.\ Phys.\  A {\bf 547}, 473 (1992).
\bibitem{protons1} L. Ray, W. R. Coker, G.W. Hoffmann, Phys. Rev. C {\bf 18}, 2641 (1978).
\bibitem{protons2} V.E. Starodubsky, N.M. Hintz, Phys. Rev. C {\bf 49}, 2118 (1994).
\bibitem{protons3} B.C. Clark, L.J. Kerr, S. Hama, Phys. Rev. C {\bf 67}, 054605 (2003). 
\bibitem{antiprotons1} A. Trzcinska \etal, Phys. Rev. Lett. {\bf 87}, 082501 (2001).
\bibitem{antiprotons2} H. Lenske, Hyperfine Interact. {\bf 194}, 277 (2009).
\bibitem{dds} T.W. Donnelly, J. Dubach, I.~Sick, Nucl. Phys.A {\bf 503}, 589 (1989).
\bibitem{couldist} C.J. Horowitz, Phys. Rev. C {\bf 57} , 3430 (1998).
\bibitem{bigprex} C.J. Horowitz, S.J. Pollock, P.A. Souder, R. Michaels, Phys. Rev. C {\bf 63}, 025501 (2001).

\bibitem{Lalazissis:1996rd}
  G.A.~Lalazissis, J.~Konig, P.~Ring,
  Phys.\ Rev.\ C {\bf 55}, 540 (1997).

\bibitem{ToddRutel:2005zz}
  B.G.~Todd-Rutel, J.~Piekarewicz,
  Phys.\ Rev.\ Lett.\  {\bf 95}, 122501 (2005).
  


\bibitem{Beiner:1974gc}
  M.~Beiner, H.~Flocard, N.~van Giai, P.~Quentin,
  Nucl.\ Phys.\  A {\bf 238}, 29 (1975).
 
\bibitem{Chabanat:1997un}
  E.~Chabanat, P.~Bonche, P.~Haensel, J.~Meyer, R.~Schaeffer,
  Nucl.\ Phys.\ A {\bf 635}, 231 (1998).
  
\bibitem{Vautherin:1971aw}
  D.~Vautherin, D.~M.~Brink,
  Phys.\ Rev.\  C {\bf 5}, 626 (1972).
  

\bibitem{alexbrown}B.A.~Brown, Phys. Rev. Lett. {\bf 85}, 5296 (2000). 
\bibitem{rNSvsRn} C.J. Horowitz, J. Piekarewicz, Phys. Rev. {\bf C64}, 062802 (2001).

\bibitem{Ozel:2010fw}
  F.~Ozel, G.~Baym, T.~Guver,
  Phys.\ Rev.\  D {\bf 82}, 101301 (2010).

\bibitem{Steiner:2010fz}
  A.~W.~Steiner, J.~M.~Lattimer, E.~F.~Brown,
  Astrophys.\ J.\  {\bf 722}, 33 (2010).

\bibitem{Hebeler:2010jx}
  K.~Hebeler, J.~M.~Lattimer, C.~J.~Pethick, A.~Schwenk,
  Phys.\ Rev.\ Lett.\  {\bf 105}, 161102 (2010).

\bibitem{MC3n} 
S. Gandolfi, J. Carlson, S. Reddy, arXiv:1101.1921.

\bibitem{isospindif} {
W. G. Lynch \etal,
arXiv:0901.0412.}

\bibitem{isospindif2}M.B. Tsang \etal, 
Phys. Rev. Lett. {\bf 102}, 122701 (2009). 
\bibitem{URCA}C.J. Horowitz, J. Piekarewicz, Phys. Rev. C {\bf 66}, 055803 (2002). 
\bibitem{cjhjp_prl}C.J. Horowitz, J. Piekarewicz, Phys. Rev. Lett. {\bf 86},  5647 (2001).
\bibitem{Reinhard}P. G. Reinhard, W. Nazarewicz, Phys. Rev. C {\bf 81}, 051303 (2010).
\bibitem{tamii} A. Tamii et al., Phys. Rev. Lett. {\bf 107}, 062502 (2011).

\bibitem{Pollock} S.~J.~Pollock, E.~N.~Fortson, and L.~Wilets,
Phys. Rev. C {\bf 46}, 2587 (1992); S.J.~Pollock and M.C.~Welliver, Phys.
Lett. B {\bf 464}, 177 (1999)

\bibitem{brownAPV} B.A. Brown, A. Derevianko, V. V. Flambaum, Phys. Rev. C {\bf 79}, 035501 (2009).
\bibitem{berkeleyAPV}
K. Tsigutkin \etal, 
  Phys.\ Rev.\  A {\bf 81}, 032114 (2010).

\bibitem{Acha:2006my}
  A.~Acha \etal 
  Phys.\ Rev.\ Lett.\  {\bf 98}, 032301 (2007).

\bibitem{Aniol:2005zf}
  K.~A.~Aniol \etal  
  Phys.\ Rev.\ Lett.\  {\bf 96}, 022003 (2006).

\bibitem{Aniol:PRC2004}
  K.~A.~Aniol \etal  
  Phys. Rev. C {\bf 69}, 065501 (2004). 

\bibitem{Alcorn:2004sb}
  J.~Alcorn  
 \etal,  
  Nucl.\ Instrum.\ Meth.\ A {\bf 522}, 294 (2004).
%

\bibitem{Sinclair2007}
C. K. Sinclair, {\it et.al.} Phys. Rev. ST Accel. Beams 10, 023501 (2007);
J. Hansknecht, {\it et.al.} Phys. Rev. ST Accel. Beams 13, 010101 (2010).

%
\bibitem{Paschke:2007zz}
  K.~D.~Paschke,
  Eur.\ Phys.\ J.\  A {\bf 32}, 549 (2007).

\bibitem{Ong} A. Ong, J. C. Berengut, V. V. Flambaum, Phys. Rev. C {\bf 82}, 014320 (2010).

\bibitem{Ban:2010wx}
  S.~Ban, C.J.~Horowitz, R.~Michaels,
J. Phys G {\bf 39} (2012) 015104.


\bibitem{RocaMaza:2011pm}
  X.~Roca-Maza, M.~Centelles, X.~Vinas, M.~Warda,
Phys. Rev. Lett. {\bf 106}, 252501 (2011).

\end{thebibliography}
